# Plasma channels under filamentation of infrared and ultraviolet double femtosecond laser pulses

A.A.Ionin,[1,*] S.I.Kudryashov,[1] D.V.Mokrousova,[1,2] L.V.Seleznev,[1]
D.V.Sinitsyn,[1] and E.S.Sunchugasheva[1,2]

[1]*P.N. Lebedev Physical Institute of the Russian Academy of Sciences, 53 Leninskiy prospect, Moscow, Russian Federation*
[2]*Moscow Institute of Physics and Technology, 9 Institutskiy Pereulok, Dolgoprudny, Moscow Region, Russia*
*Corresponding author: aion@sci.lebedev.ru

An influence of plasma channel created by a filament of focused UV or IR femtosecond laser pulse ($\lambda$=248 nm or 740 nm) on characteristics of other plasma channel formed by a femtosecond pulse at the same wavelength following the first one with varied nanosecond time delay was experimentally studied. A dependence of optical transparency of the first channel and plasma density of the second channel on the time delay was demonstrated to be quite different for such a double UV and IR femtosecond pulses.

Propagation of a femtosecond laser pulse with overcritical power through transparent medium results in self-focusing and plasma channels formation (process of filamentation [1,2]). Currently, filamentation phenomenon has wide practical application, in particular, for controlling high-voltage electric discharge [3-8] by plasma channels formed by filaments. Two sequential femtosecond pulses (i.e. a double femtosecond pulse) of infrared radiation were used to control electric discharge in [9]. An opportunity of increasing the efficiency of high-voltage discharge control with additional sequential nanosecond pulse was demonstrated in [6]. The first pulse created cold low density plasma, while the second pulse heated the plasma and increased photodetachment of electrons that reduced the breakdown discharge voltage. It was also reported in papers [8, 10, 11] that conductivity of a long plasma channel was effectively sustained by a train of UV or IR ultrashort pulses. Obviously, the efficiency of plasma channel formation and supporting electron concentration by a train of ultrashort laser pulses depends on the fact how plasma channel formed by a previous laser pulse influences upon filamentation of the subsequent pulse and its plasma channel formation. As was also shown in Ref. [12], where process of filamentation and plasma channels formation was compared for focused IR and UV ultrashort pulses, that plasma channel characteristics (shape, geometry size, plasma density, etc.) strongly depend on radiation wavelength. For the time being, studying parameters of plasma channels formed by focused double femtosecond laser pulse and comparing the results obtained for such a pulse of different (IR and UV) spectral bands has not been conducted. In this paper, an influence of plasma channel created by a filament of focused femtosecond laser pulse ($\lambda$=248 nm or 740 nm) on characteristics of other plasma channel formed by a femtosecond pulse at the same wavelength following the first one with varied nanosecond time delay was experimentally studied.

Plasma channel parameters formed by a double laser pulse consisting of two sequential ultrashort pulses were experimentally studied by using a Ti-sapphire femtosecond laser system. Laser pulses of 100 fs pulse duration (FWHM) following with 10 Hz repetition rate at the wavelengths of 248 nm and 744 nm with the laser beam radius of 4 mm ($e^{-1}$ level) were used. The laser pulse peak power was as high as a few self-focusing critical power ($P_{cr} \sim$ 3-5 GW for 744 nm, $P_{cr} \sim$ 100 MW for 248 nm [1]), which guaranteed just a single filament formation. The pulse energy was 1 mJ for the IR radiation and 35 µJ for UV one. Experimental optical scheme is shown in Fig.1. The laser beam was directed to a semitransparent mirror splitting it into two beams following along different optical passes and superposed on the next semitransparent mirror along the same optical axis. The time delay between two femtosecond pulses corresponding to the two beams was varied from 1.5 to 15 ns by changing the difference in optical passes of the pulses. The double pulse was focused by a concave mirror with 1 m focal length. The match of the optical axes for the two pulses composing the double one was controlled by CCD-camera in the far-field region. Electron density distribution of the plasma channels along the filaments formed by the double pulse was detected by using spherical electrodes. When the laser pulse went through the electrodes gap, capacity formed by the electrodes was changed by arising laser plasma, and the capacitor recharge current and corresponding voltage were recorded by an oscilloscope with resolution time ~1 ns. Since the change of the capacity depends on the electron density, length and width of the plasma channel between the electrodes, the measured current and corresponding voltage were proportional to the linear density of the plasma channel (the integral of the plasma density over the cross section of the channel). The electrodes were moved along the plasma channel giving an opportunity to measure the linear plasma distribution over the plasma channel length. First, the peak voltage corresponding to the first and second laser pulse was measured. Then the second pulse was blocked, and the signal amplitude was measured at a time corresponding to the second pulse time delay. The electron plasma density formed by the second pulse was determined through measuring the difference of the signal amplitudes. It should be noted that the voltage between the electrodes (2.5 kV) was chosen in such a way that the time of electron attachment to oxygen depending on applied voltage exceeded 15 ns [13]. Thus, the second pulse was propagated in electron-ion plasma.

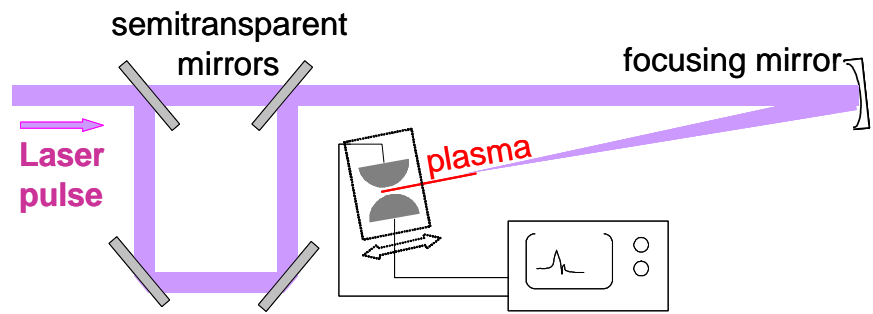

Fig. 1. Scheme of registration of the electron plasma formed by a double femtosecond pulse.

Figure 2 shows the linear density (arb.un) of the plasma channels formed under filamentation of femtosecond UV pulses versus the distance $\Delta Z = F - Z$ (Z-coordinate of the plasma channel counted off the focusing mirror along the optical axis, F - focal length) for the first and second pulses at various time delays. Propagation and focusing of the first laser pulse took place in the neutral air, and the position and shape of the plasma channel formed by the first pulse was unchanged in all measurements. At long time delays (longer than 5 ns), plasma channel formed by the second pulse was close to the form created by the first one. Probably due to refraction of the second pulse on a long plasma channel formed by the first one, a slight decrease of linear plasma density was observed near the end of the plasma channel (in front and behind the focal plane). Nevertheless, the end of the plasma channel was unchanged for different time delays. Reducing the time delay resulted in significant decrease of linear density of the plasma channel formed by the second pulse. Firstly, this decrease was assumed to be probably due to refraction of laser radiation of the second pulse on the electron-ion plasma formed by the first one, which concentration drops down with time following hyperbolic law because of electron-ion recombination within the time domain under consideration [14]. It seemed this refraction effect must have been far stronger for the infrared pulses with higher energy and a longer wavelength, creating much higher plasma concentration [12]. (It should be noted that critical plasma density for UV and IR radiation ($10^{22}$ and $10^{21}$ cm$^{-3}$, respectively) is far higher than plasma density created by ultrashort pulse for the experimental conditions in nanosecond time domain (about $10^{15}$-$10^{16}$ cm$^{-3}$). Therefore, plasma channel formed by the first laser pulse had to be transparent for the second one).

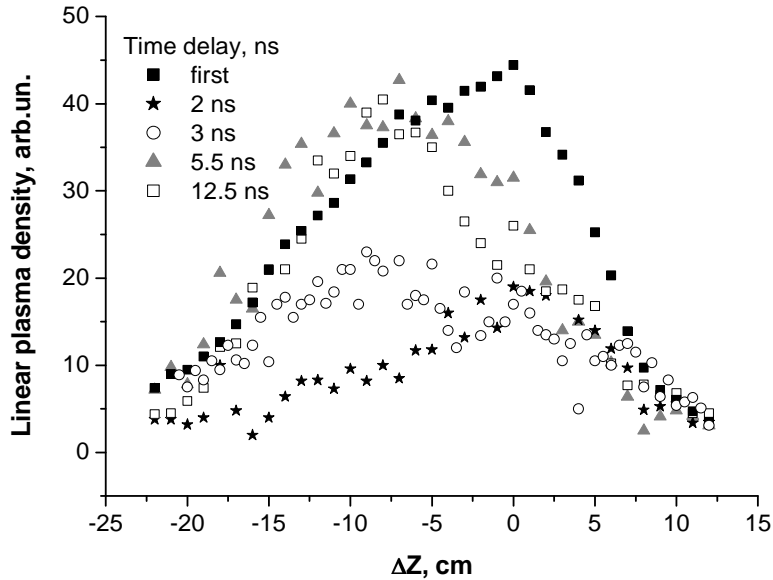

Fig.2. Linear plasma density v.s. distance $\Delta Z = F - Z$ (Z-coordinate of the plasma channel along the optical axis, F - focal length) for the first UV laser pulse and the second one for various time delays between them. Laser radiation goes from the left side.

That is why the similar experiment (under the same conditions, the optical system, and the same experimental procedure) was carried out with double pulses in the near infrared range (wavelength 740 nm). Experimental results obtained for the laser radiation with the wavelength of 740 nm are shown in Fig.3. It should be noted that at the beginning of the plasma channel the linear plasma density formed by the second pulse was slightly higher as compared to the first one. At various time delays the maximal value of linear plasma density formed by the second pulse (in the vicinity of linear focus) was approximately two times lower than plasma density formed by the first pulse. Indeed, the process of focusing and self-focusing of the second IR pulse seems to be limited by refraction on plasma created by the first pulse. However, in contrast to the UV pulses, there were no significant differences in plasma channels formed by the second IR pulse for the delay times changing from 1.6 ns to 15 ns.

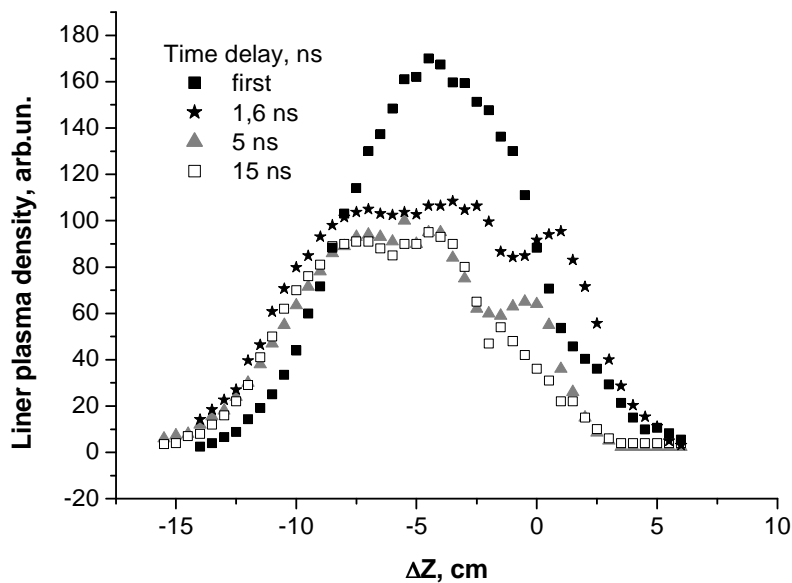

Fig. 3. Linear plasma density v.s. distance $\Delta Z = F - Z$ (Z-coordinate of the plasma channel along the optical axis, F - focal length) for the first IR laser pulse and the second one for various time delays between them. Laser radiation goes from the left side.

To determine the nature of such a difference in the experimental results for UV and IR double pulses, the energy ratio of the first and second pulses in front and behind the filamentation area was measured by a wide-aperture detector. In front of the filamentation area for various time delays, the energy ratio for the first and second both UV and IR pulses was equal one. In experiments with the wavelength of 740 nm there was no absorption at all of the second laser pulse for all time delays (from 1.5 ns to 15 ns). At 248 nm wavelength absorption of the second pulse in the plasma channel formed by previous pulse was (15 ± 3)% for the time delay of 1.5 ns. An increase of time delay resulted in a decrease of absorption: - (11 ± 3) % at 2.5 ns and (8 ± 3)% at 3.5 ns. For the time delay of 5.5 ns there was no absorption within the experimental error ranged from 0 to 3%. Any appreciable scattering of the second pulse in the plasma created by the first pulse was not observed. Thus, in the plasma induced by UV femtosecond pulse some absorption at the wavelength of 248 nm with a characteristic relaxation time of a few nanoseconds was observed, which can result in change of optical properties of the medium.

The results on UV ultrashort pulses (absorption and low plasma density at short time delay) allow one to suggest the existence of some resonance effects with a decay time of a few nanoseconds. For example, oxygen ions $O_2^+$ formed by the first laser pulse can be excited by 5 eV quantum (248 nm) resonantly from $X^2\Pi_g$ to $a^4\Pi_u$ levels [15]. It should be noted that probability of 3-quanta process of oxygen ionization is higher than 4-quanta that of nitrogen, i.e. concentration of oxygen ions is higher than nitrogen ones. The resonant absorption at the wavelength of 248 nm can lead to significant changes in the laser pulses propagation. Within a few nanoseconds due to electron-ion recombination oxygen ion concentration is reduced significantly. Properties of the medium turn to initial ones, and, as it was observed at 5.5 ns time delay, the influence of plasma created by the first pulse on the second pulse propagation becomes negligible. It should be noted that for wavelengths near 740 nm the similar transitions in oxygen ion are absent, i.e. at this wavelength such processes can not be observed, which was confirmed by experiments on the absence of absorption of the second laser pulse at time delays from 1.5 ns to 15 ns.

Thus, the propagation of focused double femtosecond pulses of UV and IR spectral range was studied. It was shown that the electron density of plasma channel formed by the second of a double IR pulse in the process of filamentation was slightly less than the density of the plasma formed by the first pulse, and did not depend on the time delay of the second pulse within the range from 1.5 to 15 ns. It was demonstrated for a double UV pulse that at time delays longer than 5 ns the linear density of plasma channel formed by the second pulse did not differ from that for the first pulse, whereas at shorter delays it fell down dramatically which can be ascribe to absorption of the second pulse in the first plasma channel.

Therefore, in the conditions of a single filament and a single plasma channel (low power excess over the critical self-focusing one) the use of a train of UV ultrashort pulses with a period less than 5 ns does not look practical for sustaining plasma in the channel. It should be noted that in case of multifilament mode (laser peak power is much higher than self-focusing one) short time delays can cause an opposite effect: the second filament and corresponding plasma channel can appear outside the filamentation zone of the previous pulse, which can lead to more uniform filling of the "thick" laser plasma channel.

We acknowledge support from the Russian Foundation for Basic Research (Grants 11-02-12061-ofi-m, 11-02-01100), EOARD Project 097007 going through ISTC Partner Project 4073 P, LPI Educational-Scientific Complex. Authors also acknowledge A.P.Napartovich, I.V.Kochetov and A.K.Kurnosov of TRINITI, and I.V.Smetanin of LPI for useful discussions.

## References

1. Couairon A., Myzyrowicz A. // Phys. Reports 441, 47 (2007)
2. Kandidov V.P., Shlenov S.A., Kosareva O.G.// Quantum Electron. **39**, 205 (2009)
3. Rambo P., Schwarz J. and Diels J.C. // J. Opt. A: Pure Appl. Opt. **3** 146–158 (2001)
4. S.L. Chin, K. Miyazaki // Jpn J. Appl. Phys. 38, 2011 (1999)

5. Pépin H., Comtois D., Vidal F., Chien C.Y. // Phys. Plasmas 8, 2532 (2001)
6. Kasparian J., Rodriquez M., Mejean G., et al. // Science 301, 61 (2003)
7. Mejean G., Ackermann R., Kasparian J. // Appl. Phys. Lett. 88, 021101 (2006)
8. Ionin A. A., Zvorykin V.D., Smetanin I.V. et al. // Appl. Phys. Lett. 100, 104105 (2012)
9. Zhe Zhang, Xin Lu, Wen-Xi Liang, Zuo-Qiang Hao, Mu-Lin Zhou, Zhao-Hua Wang, Xun Liu and Jie Zhang // Optics Express **17** (5) 3461 (2009)
10. V D Zvorykin, A A Ionin, A O Levchenko, G A Mesyats et al // Quantum Electron **43** (4), 339 (2013)
11. Xiao-Long Liu, Xin Lu, Jing-Long Ma, Liu-Bin Feng et al // Optics Express **20** (6) 5968 (2012)
12. A A Dergachev, A A Ionin, V P Kandidov, L V Seleznev et al // Quantum Electron **43** (1) 29 (2013)
13. A.V.Shutov, I.V.Smetanin, A. A. Ionin, A.O.Levchenko, L.V.Seleznev, D.A.Sinitsyn, N.N.Ustinovskii, and V.D.Zvorykin // Appl. Phys. Lett. 103, 034106 (2013)
14. Raizer Yu.P. // Physics of gas discharge. Moscow. Nauka (1987) (in Russian)
15. Krupenie P.H. // J.Phys.Chem Ref.Data, **1** (2) 423 (1972)